\documentclass[a4paper,11pt]{article}
\usepackage{pos}

\usepackage{xspace}
\newcommand{\Xmax}{X$_\text{max}$\xspace}
\newcommand{\aboot}{$\sim$}

\usepackage{cleveref}
\usepackage{wrapfig}

\newcommand{\shrink}{\vspace{-0.3cm}}

\newlength{\bibitemsep}
\newlength{\bibparskip}
\let\oldthebibliography\thebibliography
\renewcommand\thebibliography[1]{%
  \oldthebibliography{#1}%
  \setlength{\parskip}{\bibitemsep}%
  \setlength{\itemsep}{\bibparskip}%
}

\usepackage{subfiles} % Best loaded last in the preamble

\title{Simulation Study of the Observed Radio Emission of Air Showers by the IceTop Surface Extension}
 \ShortTitle{Simulation Study of Radio Emission at IceTop}

\author{The IceCube Collaboration \\{\normalsize \normalfont(a complete list of authors can be found at the end of the proceedings)}}

\emailAdd{alanc@udel.edu}

\abstract{Multi-detector observations of individual air showers are critical to make significant progress to precisely determine cosmic-ray quantities such as mass and energy of individual events and thus bring us a step forward in answering the open questions in cosmic-ray physics. An enhancement of IceTop, the surface array of the IceCube Neutrino Observatory, is currently underway and includes adding antennas and scintillators to the existing array of ice-Cherenkov tanks. The radio component will improve the characterization of the primary particles by providing an estimation of X$_\text{max}$ and a direct sampling of the electromagnetic cascade, both important for per-event mass classification. A prototype station has been operated at the South Pole and has observed showers, simultaneously, with the tanks, scintillator panels, and antennas. The observed radio signals of these events are unique as they are measured in the 70 to 350\,MHz band, higher than many other cosmic-ray experiments. We present a comparison of the detected events with the waveforms from CoREAS simulations, convoluted with the end-to-end electronics response, as a verification of the analysis chain. Using the detector response and the measurements of the prototype station as input, we update a Monte-Carlo-based study on the potential of the enhanced surface array for the hybrid detection of air showers by scintillators and radio antennas.

\vspace{4mm}
{\bfseries Corresponding author:}
Alan Coleman$^{1*}$\\
{$^{1}$ \itshape Bartol Research Institute and University of Delaware Department of Physics \& Astronomy,\\
\indent\hspace{0.3cm}Newark, DE, USA}\\[4mm]
$^*$ Presenter
\FullConference{37$^{\rm{th}}$ International Cosmic Ray Conference (ICRC 2021)\\
		July 12th -- 23rd, 2021\\
		Online -- Berlin, Germany}
}

\begin{document}
\maketitle

\section{CR Detection Using Surface-Radio at the South Pole}

For the next generation of cosmic ray (CR) arrays, a more accurate identification of the properties of the primary particle is a driving concern for the design of detector components.
In the field of high-energy CRs, there are still correlated unknowns such as the origins and acceleration mechanisms.
A major challenge to making progress towards these open questions is the underlying CR mass distributions.
The second unknown is an accurate description of the hadronic interactions that govern particle generation at center-of-mass energies above those measured at the LHC.
Current experiments have shown that above CR energies of \aboot30\,PeV, modern hadronic interaction models do not accurately reproduce the observed distribution of muons~\cite{soldin:2021icrc}.
One method towards making progress in understanding these linked topics is to disentangle the distributions of air shower particles, specifically the electromagnetic (EM) and muonic content.

Radio antennas have reached maturity in their use to detect air showers and are only sensitive to the EM content of the shower.
The ability to determine the EM energy and \Xmax has already been demonstrated by experiments such as the Pierre Auger Observatory~\cite{aab2016energy}, LOFAR~\cite{Buitink:2014eqa}, and Tunka-Rex~\cite{bezyazeekov2018reconstruction} to an accuracy of better than 20\% and 30\,g\,cm$^{-2}$, respectively.
These quantities are important when trying to classify the primary mass of a given air shower observation.

IceTop is a 1\,km$^2$ CR detector and is part of the IceCube Neutrino Observatory, located at the South Pole.
% and is designed an important too to study CR and particle physics while at the same time acting as a veto for the in-ice neutrino detector.
It currently consists of 162 ice Cherenkov tanks which detect the emission of relativistic particles that enter their volume.
The IceTop enhancement foresees the addition of scintillator panels and radio antennas to the current footprint to extend the CR and neutrino program of the Observatory~\cite{schroder2019science}.
In this work, we detail the end-to-end simulation chain of the antennas in \cref{s:SimSoftware}, present a calculation of the expected sensitivity of the final array in \cref{ss:ExpectedEventDistribution}, and show a comparison of simulated and observed waveforms using the currently deployed prototype station~\cite{roxmarieProceeding,hrvojeProceeding} in \cref{s:ObservedEvents}.

\section{Simulation and Analysis Software}
\label{s:SimSoftware}

The suite of analysis software for the processing of observed and simulated radio waveforms is included in the larger IceCube framework, IceTray~\cite{de2005icetray}. 
The additional radio-specific software includes a repository of modules which are used to analyze waveforms in the time and frequency domains, (de)convolve the hardware responses, perform frequency filtering (see \cref{ss:FrequencyFilters}), and calculate standard physics quantities such as the radio-frequency energy fluence.

For the results shown in this work, the air-shower Monte Carlo (MC) package, CORSIKA, and its radio emission extension, CoREAS, were used to calculate both the particle emission and electric field waveforms~\cite{corsika} for the proposed IceTop Enhancement layout. 

The injection of the CR air-shower secondaries into the particle detectors, the resulting light yield, and the photo-multiplier response are handled with a dedicated Geant4 simulation on a per-event basis~\cite{agnieszka2019scintsim}.
The fidelity of the scintillator injection code is particularly important as the panels provide the external trigger for the readout of the antennas and are therefore an integral part of simulating air shower events with antenna waveforms.

% An essential requirement of the radio enhancement is an accurate determination of the \emph{absolute} amplitude of the impinging electric field.
% The accuracy to which this is measured linearly translates to the systematic uncertainty in the EM energy content of an air shower.
% Thus, a detailed simulation of all the detector components are required to compare observed signals to those from MC.

The simulation of the radio response begins with the output of the CoREAS package, the electric field as a function of time, $\vec{E}(t)$.
The beginning of the waveform is first pre-padded with $\vec{0}$ such that it is 5000 bins long and then resampled from the original 0.2\,ns time-step binning to 1\,ns, producing a waveform that is 1\,$\mu$s long.
The far-field antenna response of the SKALA v2 crossed-dipole was simulated for all Poynting vector directions in steps of 1$^\circ$ and in frequencies in 1\,MHz steps for 50 - 350\,MHz~\cite{SKALAV2_LNA}.
The simulated vector effective length, $\vec{\mathcal{L}}(\Theta,\Phi)$, where $\Theta$ and $\Phi$ define the propagation direction of the far-field wave front in spherical coordinates, is used to calculate the voltage in the antenna, $V(f) = \vec{\mathcal{L}}(\Theta,\Phi;f) \cdot \vec{E}(f)$.
% The values of $\Theta$ and $\Phi$ are approximated as those of the primary CR's momentum vector, shown to be valid to within $\simeq$1$^\circ$~\cite{}.
% As needed, the values of $\vec{\mathcal{L}}$ are calculated via a cubic interpolation using the tabulated values in $f$, $\Theta$, and $\Phi$.

%
%
\begin{figure}
    \shrink
    \centering
    \includegraphics[width=0.99\textwidth]{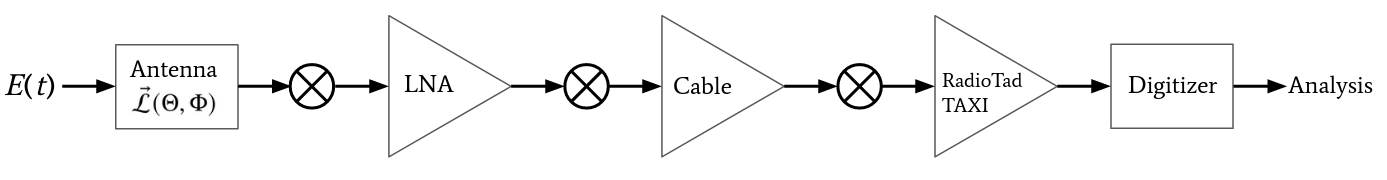}
    \shrink
    \caption{A schematic of the simulation chain is shown above. The voltage in the antenna is calculated using a simulated model of the vector effective length. Convolutions are used to account for the various hardware components after which there is a final digitization to ADC units.}
    \shrink
    \label{fig:ConvolutionSchematic}
\end{figure}
\begin{wrapfigure}{r}{0.5\textwidth}
  \shrink
  \begin{center}
    \includegraphics[width=0.48\textwidth]{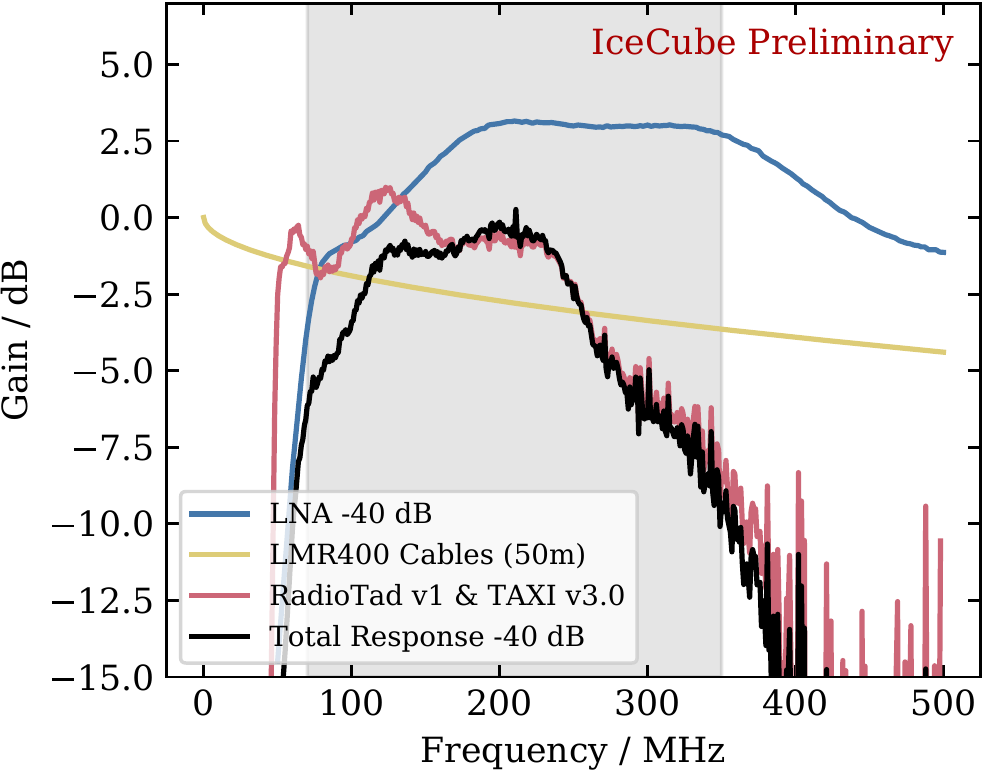}
  \end{center}
  \shrink
  \caption{The power gain of the various components is shown above. The gray band indicates the nominal frequency range of the surface array, 70 - 350\,MHz.}

    \label{fig:ElectronicsResponse}
\end{wrapfigure}
The voltages that are produced in the SKALA antenna are then folded with the response of the hardware readout chain, including a low-noise amplifier (LNA), coaxial cables, and a pre-processing board.
All components are assumed to be working in their linear regime and thus can be directly combined as shown in \cref{fig:ConvolutionSchematic}.
The gain of the LNA, mounted to the top of each SKALA antenna, has been simulated and provides an approximate $+$40\,dB amplification.
Next, the 50\,m LMR400 coaxial cables, the response of which has been measured as a function frequency and temperature, is $-$2 to $-$1\,dB, with higher attenuation at high frequency.
Finally the voltage is folded with the DAQ which includes the signal pre-processing board, RadioTad, and an ADC converter, TAXI v3.0~\cite{roxmarieProceeding}.
The combined response of the DAQ has been measured in the lab and is $-$10 to $+$1\,dB, varying over the frequency band of interest.
The frequency-dependent responses of the components described above are shown in \cref{fig:ElectronicsResponse}.
The final step includes a digitization of the signal with 14 bit precision and a dynamic range of 1\,V.
The combined response of this signal chain ensures no saturation for CR energies up to $10^{18}$\,eV.
\section{Detection of CR Events Using Surface Radio}
\label{s:DetectionOfCR}

The sensitivity of the antenna array to air showers is determined by the density of the antennas, the external trigger of the scintillator panels, and the characteristics of the background noise.
The array will include 92 antennas distributed over the current IceTop footprint.
% For vertical air shower trajectories, the scintillator panels will trigger with $>$98\% efficiency on primary energies of $\simeq$300\,PeV for XXX$^\circ$ to YYY$^\circ$ and $\simeq$3\,PeV for XXX to YYY$^\circ$.
% Above XXX$^\circ$, the projected area of the XXX\,cm thick panels is quite small and are thus a relevant constraint on air shower detection with radio, with a respective triggering threshold of XX\,PeV.
% The background noise has been measured at the pole using a prototype station~\cite{} consisting of a single station of 8 scintillator panels and 3 antennas.
The primary background is a result of the diffuse flux emitted from Galactic and extra-Galactic sources with additional spikes caused by terrestrial radio-frequency interference (RFI).
This section will discuss a method to weight the individual signal frequencies, a calculation of the sensitivity of the antenna array, and a comparison of observed and simulated waveforms.

\subsection{Post-processing Frequency Filters}
\label{ss:FrequencyFilters}
\begin{figure}
    \shrink
    \centering
    \includegraphics[width=0.65\textwidth]{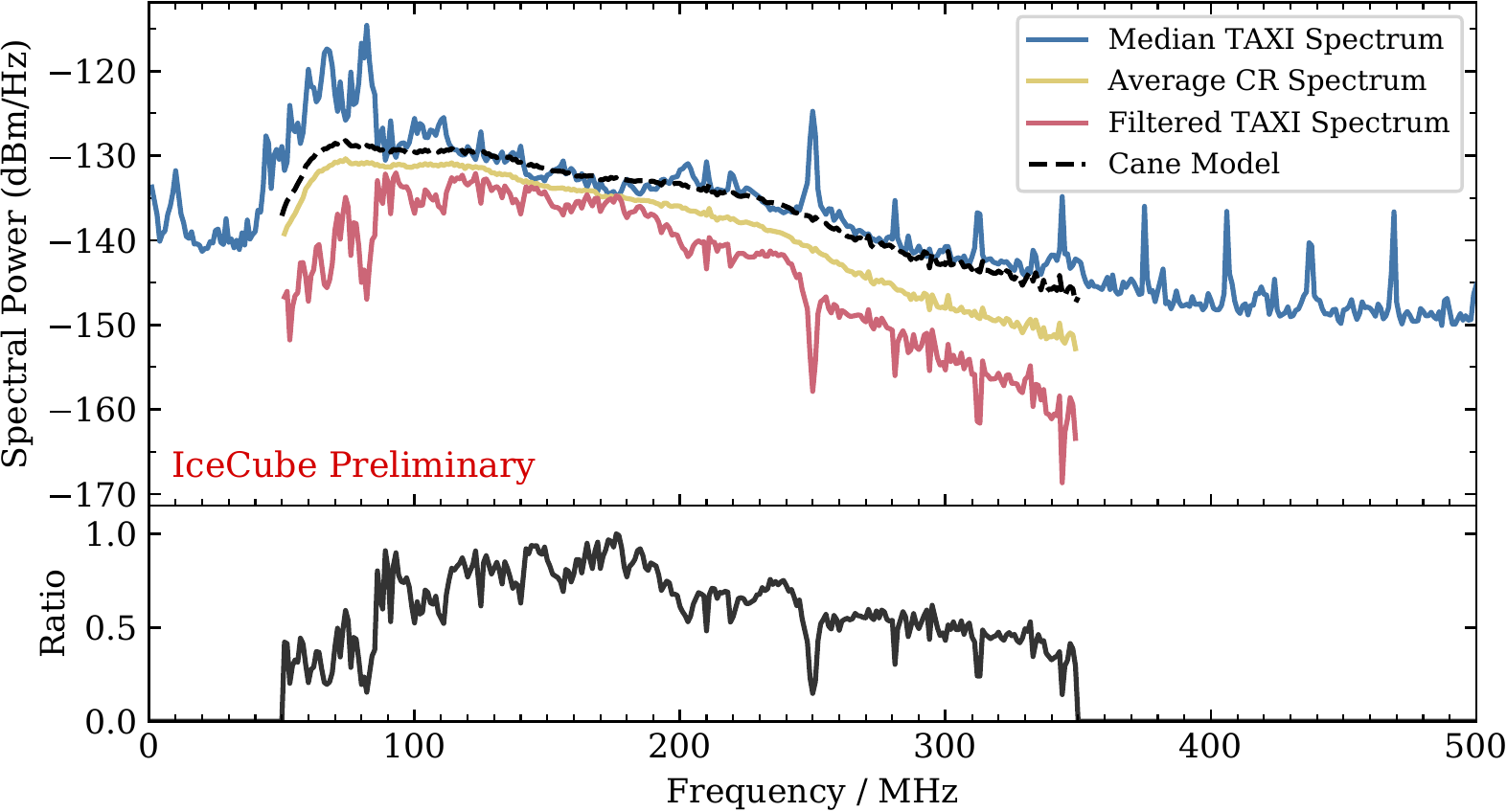}
    \caption{The median spectral power of the local background as measured by the prototype station at the South Pole using TAXI v3.0~\cite{roxmarieProceeding} is shown in blue. The Cane model of Galactic and extra-Galactic diffuse radio emission, folded with the detector response, is shown in dashed black. The average CR emission spectrum and the resultant frequency weighting are shown in gold and red, respectively. The bottom plot directly shows the frequency weights, \cref{eq:CRFilter}, directly.}
    \shrink
    \label{fig:SpikeFilter}
\end{figure}
%
%

% The radio-frequency background can be conceptually separated into narrow- and broad-band classes.
% Narrow-band noise has been observed at the pole 
% The broad-band noise, for instance, the diffuse Galactic emission, while potentially coherent and board-band at the source, is non-coherent when summed over all possible arrival directions at a given antenna.
% Coherent background, inherently broadband or multi-band, is more problematic as it can result in pulses which are non-distinguishable from air-shower pulses.

The spectral power from background noise has been characterized using measurements from the prototype station at the South Pole deployed in 2020, which includes the radioTad v1 and TAXI v3.0 readout system~\cite{roxmarieProceeding} and is shown in \cref{fig:SpikeFilter} (blue).
The distribution is the bin-wise median of 10k 1$\mu$s waveforms.
For comparison, the Cane model~\cite{cane1979spectra} of Galactic and extra-Galactic diffuse radio emission has been folded with the antenna and detector responses as described in \cref{s:SimSoftware} (dashed black).
Above 100\,MHz, the observed power is consistent with the Cane model with additional RFI peaks (ex: at 250\,MHz).
Below 100\,MHz, the prototype version of TAXI emits power which is 10\,dB above the Cane model.
This is generated by the DC-DC converters and has been mitigated in the improved version (TAXI v3.2) that will be deployed in the future Pole seasons~\cite{roxmarieProceeding}.
Other RFI peaks are evident at the highest energies but are above the nominal band where detection is limited by $-$150\,dBm/Hz thermal noise.

Standard methods exist for reducing coherent noise such as applying a median filter to individually measured waveforms, wherein the amplitudes of individual frequency bins, $a_i$, are replaced with the median value in a window of half-width, $hw$, $a_i' = {\rm Median}(a_{i-hw}, ..., a_{i+hw})$.
Another typical method involves applying hardware- or software-based notch filters wherein specific, narrow frequency bands are removed.
We build on these two ideas to create a frequency weighting scheme which notches the signals by using the median background power spectrum without the need to choose which frequencies to notch.

For this work, a library of simulated air showers was created which included discrete zenith, azimuth, and energy bins.
The directions were chosen based on the direction of the local geomagnetic field which is 17.8$^\circ$ from the zenith and 30.7$^\circ$ west-of-north at the South Pole.
Zenith angles were chosen in 17$^\circ$ steps from 0$^\circ$ to 68$^\circ$ and azimuth angles were chosen to maximize and minimize the angle with the geomagnetic field, 30.7$^\circ$ west-of-north and east-of-south.
% As the radio emission is mostly dependent on the distance of the observer from \Xmax and with an amplitude that scales with the number of EM particles, 
Proton and iron primaries were taken as limiting cases, using Sibyll 2.3d as a hadronic-interaction model~\cite{Engel:2019dsg}.
The April South Pole atmosphere was used.
The radio emission at each antenna was propagated through the simulation chain described in \cref{s:SimSoftware}. 

To construct the frequency weights, we begin the with bin-wise median amplitudes, $m_i$, of the measured background as shown in blue in \cref{fig:SpikeFilter}.
We then calculated the average emission from 300\,PeV air showers as a function of frequency, $\mathcal{C}_i$, using only the three antennas per event with the largest Hilbert peak.
Finally the weighting scheme is given by the ratio of these two quantities,
\begin{equation}
    w_i = \mathcal{N}(\mathcal{C}_{i} / m_{i})^p.
    \label{eq:CRFilter}
\end{equation}
The parameter, $p$, can be set by the user to increase or decrease the weighting.
Note that for $p=1$, the application of the weights, $w_i$, will simply change $a_i \longrightarrow \mathcal{C}_{i}$, on average, and that for this paper, $p=2$, has been used.
A normalization factor, $\mathcal{N}$, is used such that the maximum value of $w_i$ is 1.
The distributions of $\mathcal{C}_i$ and $m_{i}\times w_i$ are shown in gold and red, respectively in \cref{fig:SpikeFilter}.
The bottom plot shows the ratio from which the weights were constructed (black).

This weighting scheme, $w_i$, goes beyond a simple filtering and additionally accentuates frequencies where the typical air shower spectral shape is harder than that of the background, $\simeq$100 to 200\,MHz.
This process also inherently applies a notch filtering where the prominence of each RFI spike becomes an equally deep notch.

\subsection{Expected Observed Events at the Pole}
\label{ss:ExpectedEventDistribution}
    
The readout of the antenna waveforms will be initiated by the scintillator panels and thus all events will have corresponding waveforms.
The decision to include radio information in a given antenna during an analysis relies on the ability to determine meaningful properties of the air shower pulse amongst all sources of background noise.
For this, we use a basic criteria of a signal-to-noise-ratio (SNR) threshold cut as defined by the square of the ratio of the peak amplitude of the Hilbert envelope and the RMS of the voltage of the noise,
\begin{equation}
    {\rm SNR} = \left( \frac{{\rm Hilbert Peak}}{{\rm RMS}} \right)^2.
\end{equation}
To reduce contamination of the true signal when calculating the RMS, we use separate signal and noise windows, 200\,ns and 400\,ns long, respectively.
The size of the signal window is motivated by the $\leq 1^\circ$ angular resolution of the scintillator reconstruction~\cite{agnieszka2019scintsim} and the $\leq$\,300\,m radius of the Cherenkov ring for trajectories with $\theta < 65^\circ$.

\begin{figure}
    \shrink
    \centering
    \includegraphics[width=0.45\textwidth]{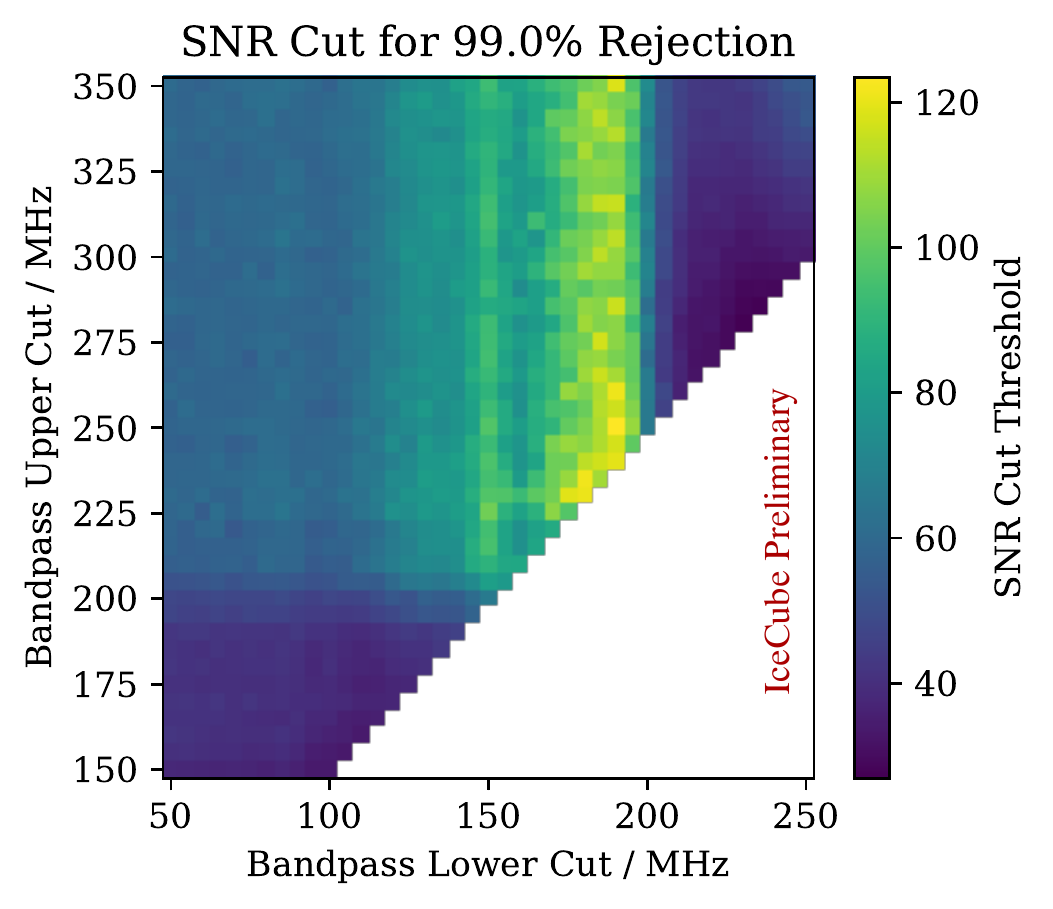}
    \hspace{1cm}
    \includegraphics[width=0.435\textwidth]{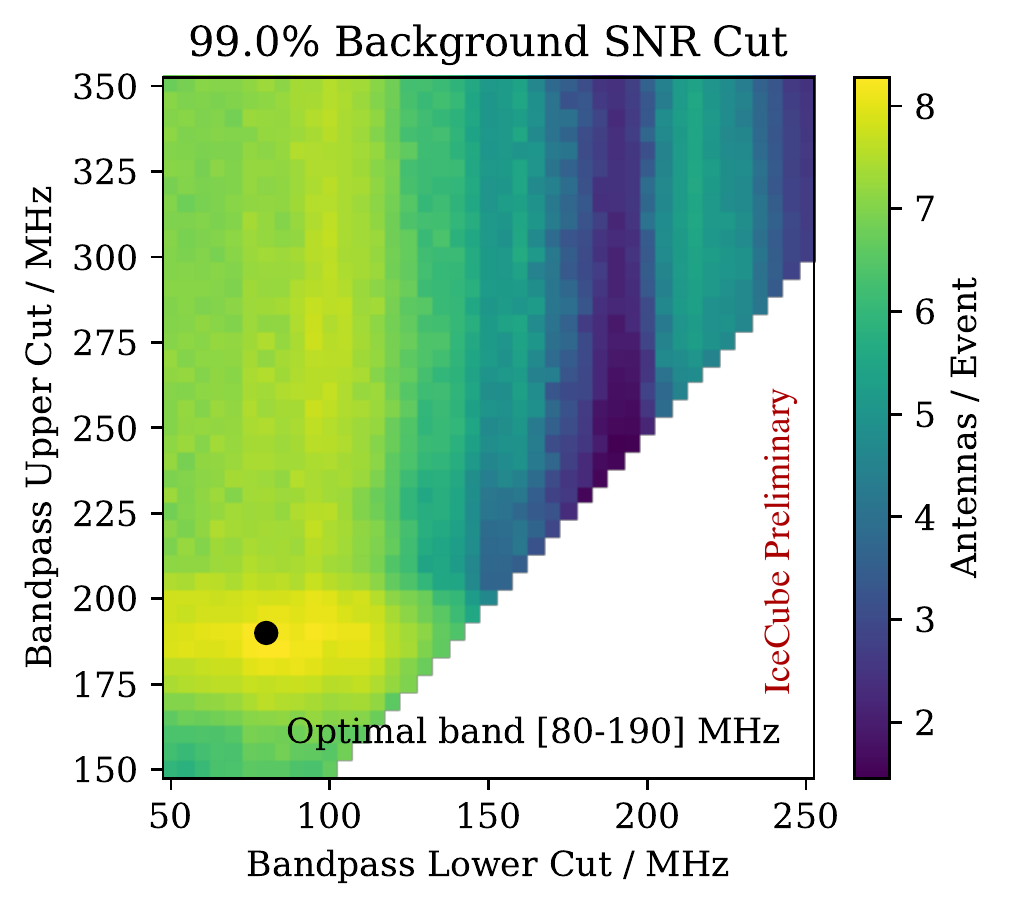}
    \caption{Left: The SNR cut values required to reject 99\% of background waveforms as a function of the low- and high-pass frequencies. Right: The number of antennas per event which pass the SNR cut for the bandpass limits. The Black dot indicates the optimal band (corresponding to a SNR threshold of 41.4).}
    \shrink
    \label{fig:SNRCutValues}
\end{figure}

The SNR threshold was then determined by using background waveforms measured at the Pole.
Again 10k waveforms were analyzed using randomly positioned signal and noise windows with the constraint that the noise window is at least 50\,ns after the signal window.
The frequency weights, \cref{eq:CRFilter} were applied to the waveforms and the SNR values were calculated.
Since the optional frequency band to use for determining detection is a priori unknown, a final bandpass filter was applied for various high-pass (50 to 250\,MHz) and low-pass (150 to 350\,MHz) values.
The SNR cut values which exclude 99\% of background waveforms are shown in \cref{fig:SNRCutValues} as a function of the considered band.

The optimal frequency band to distinguish background from CR pulses was then determined using the library of showers, described above.
First, showers with energies of 300\,PeV were processed through the simulation chain.
Background waveforms were added to the simulations directly and the frequency weights were applied.
The signal window was centered on the location of the underlying peak, known from the scintillator reconstruction.
The range of possible bandpasses as shown on the left in \cref{fig:SNRCutValues} was applied and the measured SNR value was checked against the corresponding threshold.
The optimal band was then taken to be that which maximized the average number of antennas per event that passed the SNR cut.
The detection efficiency (see below) was calculated and the above procedure was again repeated but instead using all the direction and energy bins for which the array has a 30\% to 70\% chance to trigger.
The optimal band was found to be 80 to 190\,MHz, as shown on the right in \cref{fig:SNRCutValues}, a band consistent to a previous study which used only ideal noise~\cite{aswathi2021photons}.

\begin{figure}
    \centering
    \includegraphics[width=0.42\textwidth]{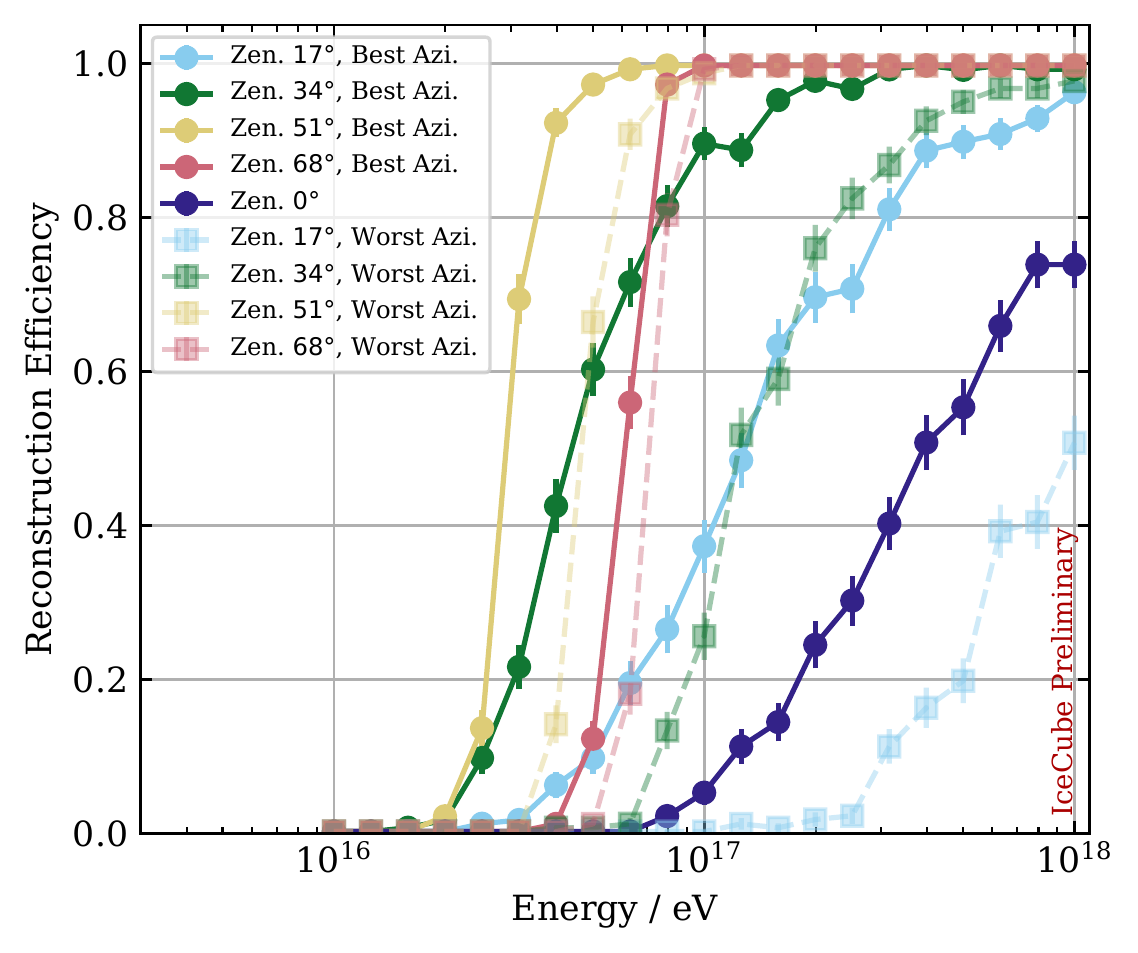}
    \hspace{0.7cm}
    \includegraphics[width=0.42\textwidth]{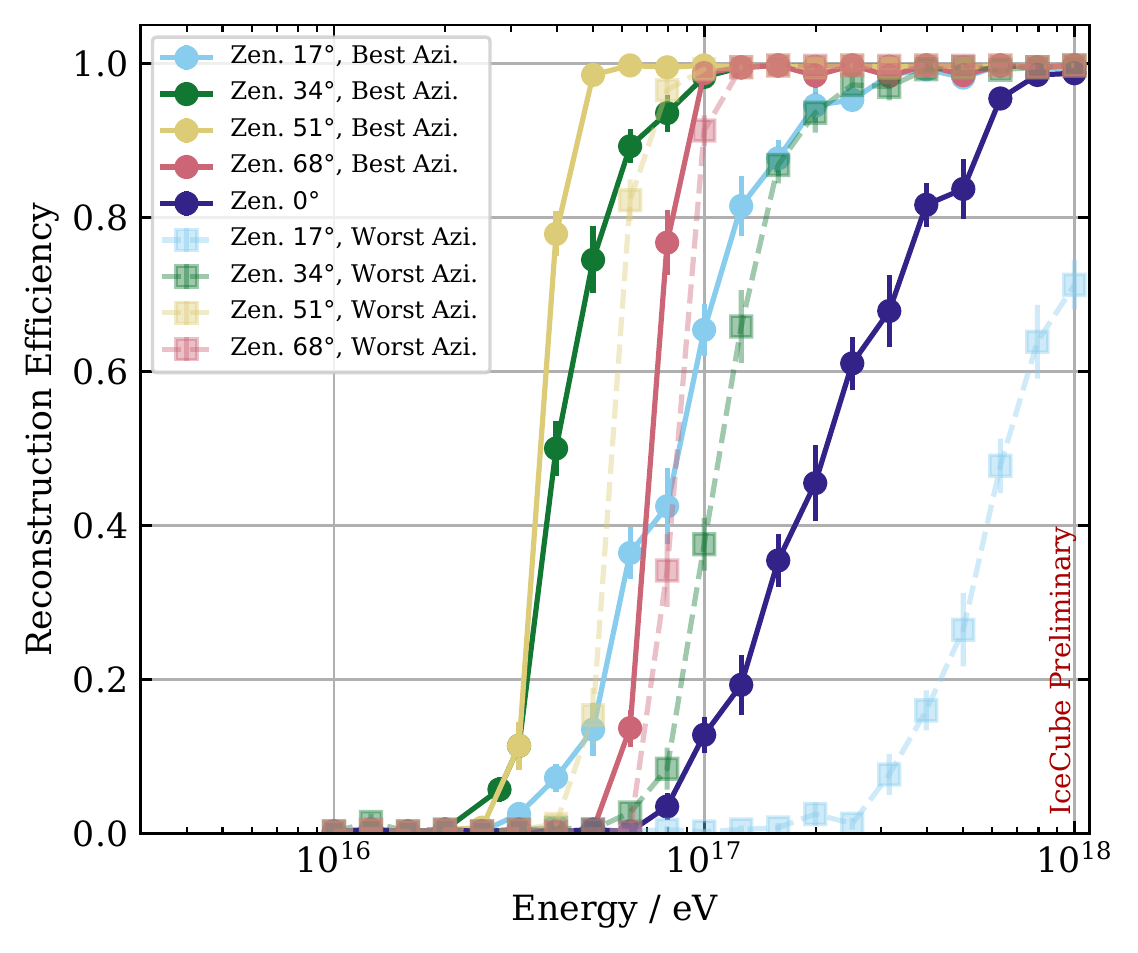}
    \caption{The fraction of simulated CoREAS events which can be reconstructed using a plane wave to within 5$^\circ$ of the MC truth are shown for proton (left) and iron (right) primaries. Background waveforms from the South Pole, recorded with TAXI v3.0 DAQ, were used. The selection of antennas to use in the reconstruction includes an SNR cut that rejects 99\% of background pulses.}
    \shrink
    \label{fig:Efficiency}
\end{figure}

Using the library of air showers, the reconstruction efficiency was determined using the full simulation chain including the addition of background noise waveforms, frequency-weighting scheme, and optimal band, as defined above.
Further, the scintillator trigger algorithm was applied wherein stations were individually triggered with the requirement that at least 3 stations recorded $\geq0.5$\,MIP (minimum-ionizing particle).
Using the antennas in the triggered stations which also passed the SNR cut, a plane front reconstruction was performed.
If the direction of the plane wave and the MC truth agreed to within 5$^\circ$, the event was considered successfully reconstructed.
The reconstruction efficiency for proton and iron primaries is shown in \cref{fig:Efficiency} as a function of energy and zenith angle for the best- and worst-case azimuth angles.
The reconstruction efficiency for proton is better for high angles, consistent with the increased EM content compared to heavier primaries.
However, for more vertical trajectories, a better efficiency is seen for the iron primaries.
The yearly average atmospheric overburden at the South Pole is about 690\,g/cm$^{2}$ which is approximately the ${\langle X_{\rm max} \rangle}$ of a vertical 100\,PeV proton shower.
Thus, for more vertical protons at-and-above this energy, much of the radio emission is being truncated by the ground and not radiated to the antennas.
Additionally, the $\simeq$1$^\circ$ Cherenkov opening angle is not wide enough to produce a distinct ring on the ground and thus the footprint is thus relatively small for zenith angles less than 40$^\circ$.
For 68$^\circ$ showers, the reconstruction efficiency is limited by the sensitivity of the scintillators to provide an external trigger.

\subsection{Comparison to Observed Events}
\label{s:ObservedEvents}
A number of events have already been observed using the prototype station at the Pole.
The timestamp of events that were triggered by the scintillator array have been cross checked with the IceTop tanks as a confirmation that the array is triggering on air showers~\cite{hrvojeProceeding}.
Both the IceTop and scintillator detectors are sensitive to air showers with energies at and below 1\,PeV and will thus provide simultaneous estimates of the air-shower properties.
Above 30\,PeV, the IceTop and scintillator reconstruction accuracy is $\sigma_{\rm core}\leq10$\,m, $\psi_{\rm direction}\leq1^\circ$, and $\sigma_E/E \leq 15\%$. 
% Thus each shower for which an antenna-based reconstruction is expected will include a simultaneous reconstruction by the other two surface detectors, each with independent estimations of the core ($\leq10$\,m), arrival direction ($\leq1\degr$), and energy ($\leq15\%)$.

\begin{figure}
    \centering
    \includegraphics[width=0.9\textwidth]{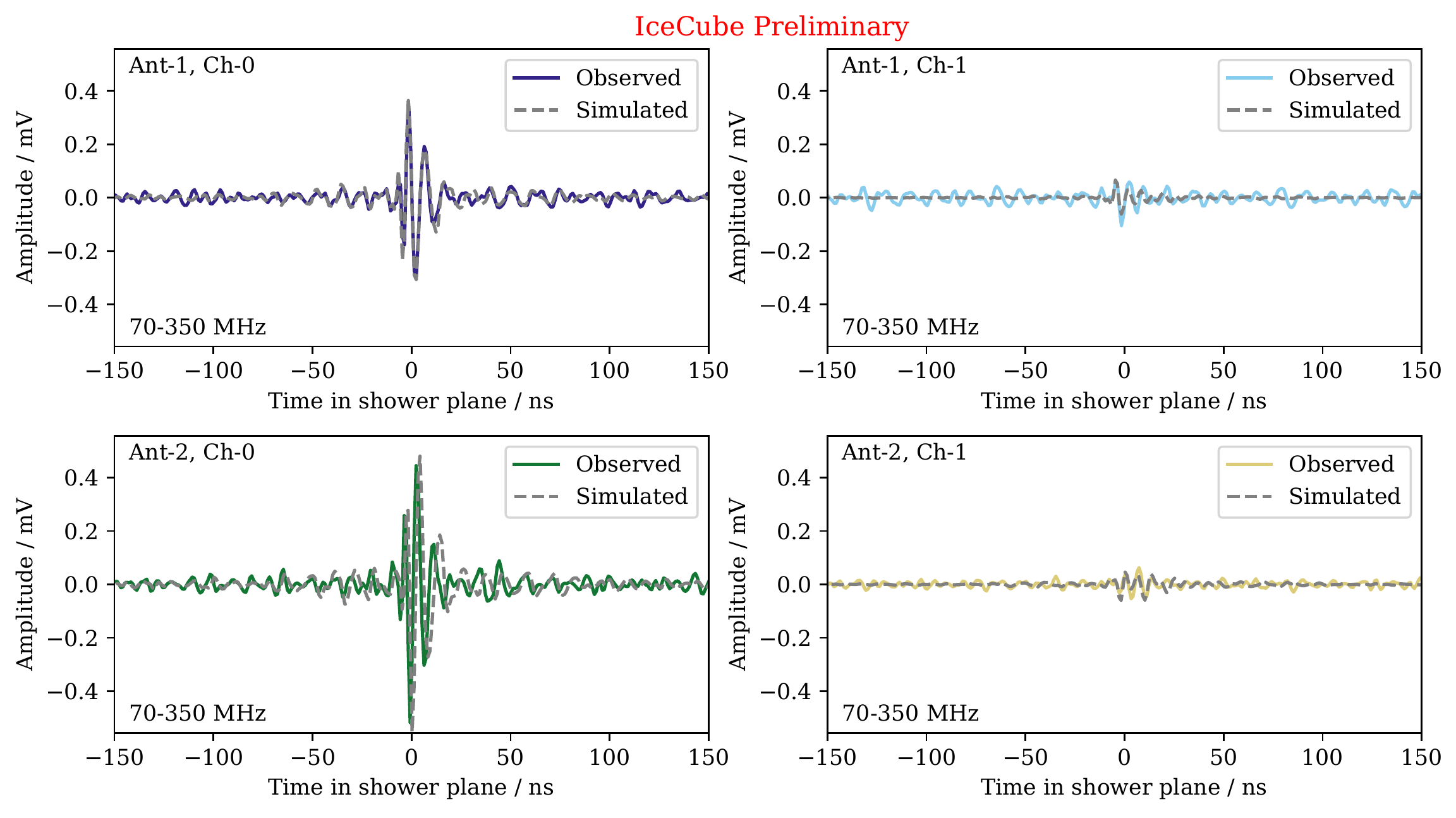}
    \caption{Voltage waveforms from an observed event (solid color) compared to the waveforms from a CoREAS simulation (dashed gray) with initial conditions as given by the IceTop reconstruction ($\theta = 32^\circ$, $\phi = 275^\circ, E_{\rm IT}=240$\,PeV).}
    \label{fig:RealEvent}
    \shrink
\end{figure}

As a last cross check of the radio simulation chain, air showers consistent with the IceTop reconstruction were created.
The radio emission was propagated into the three antennas of the prototype station and the frequency weighting scheme was applied.
An example of the simulated and observed waveforms of a 32$^\circ$, 240\,PeV air shower is shown in \cref{fig:RealEvent}.
Qualitatively, the simulated waveforms look to be consistent with that of the expected ones in phase and absolute amplitude, giving further confidence that the antennas are indeed recording air shower pulses.

Further systematic studies to determine consistency with the expected distribution of flux are ongoing~\cite{hrvojeProceeding}.
So far the arrival distribution, corresponding estimated energies from the IceTop reconstruction, and core distributions show agreement.
\section{Conclusion}
The radio antennas will be an important component to the enhanced IceTop array.
With the ability to determine the electromagnetic energy and the depth of shower maximum of CR air showers, future analyses will have increased mass discrimination power.
In this work we detailed a method to weight individual frequencies to remove RFI spikes.
Using this technique, the efficiency to reconstruct air showers using at least 3 radio antennas was determined.
For the most vertical showers, $\leq$\,$17^\circ$, reconstructions using the antennas are only possible above 100\,PeV.
For more inclined showers, up to 50$^\circ$, the energy threshold is as low as 30\,PeV, with strong azimuthal dependence.
For the most inclined showers studied here, 68$^\circ$, a reconstruction using the antennas is limited by the external scintillator trigger.

Further improvements to enhance the sensitivity of the antennas to air showers are being explored including machine learning techniques to remove noise from the waveforms directly~\cite{abdulProceeding, bezyazeekov2021reconstruction}.
Even for the current detection thresholds, an increase in selected antennas can already provide a large boost as other experiments have shown that five or more antennas are needed to reconstruct \Xmax with precision comparable to optical techniques.

{\small \medskip \noindent\textbf{Acknowledgement:} U.S. National Science Foundation-EPSCoR (RII Track-2 FEC, award ID 2019597)}
\vspace{-0.25cm}
% \begin{thebibliography}{99}
\bibliographystyle{ICRC}
\bibliography{biblio}
% \bibliographystyle{unsrt}
% \def\vyp#1#2#3{\textbf{#1} (#2) #3}
% \footnotesize
% \raggedright
% \setlength{\parskip}{0ex}

% \end{thebibliography}

\clearpage
\section*{Full Author List: IceCube Collaboration}

\scriptsize
\scriptsize
\noindent
R. Abbasi$^{17}$,
M. Ackermann$^{59}$,
J. Adams$^{18}$,
J. A. Aguilar$^{12}$,
M. Ahlers$^{22}$,
M. Ahrens$^{50}$,
C. Alispach$^{28}$,
A. A. Alves Jr.$^{31}$,
N. M. Amin$^{42}$,
R. An$^{14}$
% ...
\\

\noindent
% ...
$^{12}$ Universit{\'e} Libre de Bruxelles, Science Faculty CP230, B-1050 Brussels, Belgium \\
% ...
$^{14}$ Department of Physics and Laboratory for Particle Physics and Cosmology, Harvard University, Cambridge, MA 02138, USA \\
% ...
$^{17}$ Department of Physics, Loyola University Chicago, Chicago, IL 60660, USA \\
$^{18}$ Dept. of Physics and Astronomy, University of Canterbury, Private Bag 4800, Christchurch, New Zealand \\
% ...
$^{22}$ Niels Bohr Institute, University of Copenhagen, DK-2100 Copenhagen, Denmark \\
% ...
$^{28}$ D{\'e}partement de physique nucl{\'e}aire et corpusculaire, Universit{\'e} de Gen{\`e}ve, CH-1211 Gen{\`e}ve, Switzerland \\
% ...
$^{31}$ Karlsruhe Institute of Technology, Institute for Astroparticle Physics, D-76021 Karlsruhe, Germany  \\
% ...
$^{42}$ Bartol Research Institute and Dept. of Physics and Astronomy, University of Delaware, Newark, DE 19716, USA \\
% ...
$^{50}$ Oskar Klein Centre and Dept. of Physics, Stockholm University, SE-10691 Stockholm, Sweden \\
% ...
$^{59}$ DESY, D-15738 Zeuthen, Germany \\
% ...

\normalsize

% \clearpage
% \input{ICRC2021-SurfaceRadioSimulation/ExecSummary}

\end{document}